\documentclass[pra, showpacs, twocolumn, floatfix]{revtex4-1}

\usepackage{graphicx}
\usepackage{amsmath, amsfonts, amssymb, bm}

\makeatletter
\let\@bib@Xauthor\@undefined
\makeatother

\renewcommand{\vec}[1]{\boldsymbol{#1}}

\begin{document}

\title{Kapitza-Dirac scattering of electrons from a bichromatic standing laser wave}

\author{Matthias M. Dellweg}
\author{Carsten M\"uller}
\affiliation{Institut f\"ur Theoretische Physik I, Heinrich-Heine-Universit\"at D\"usseldorf, Universit\"atsstr. 1, 40225 D\"usseldorf, Germany}

\date{\today}

\begin{abstract}
Coherent scattering of an electron beam by the Kapitza-Dirac effect from a standing laser wave which
comprises two frequency components is studied. To this end, the Schr\"odinger equation is solved
numerically with a suitable ponderomotive potential. Besides, an analytical solution for electron
diffraction in the asymptotic domain of large field amplitudes is obtained and a mathematical model
in reduced dimensionality for the scattering amplitude in the Bragg regime is presented. We demonstrate
distinct interference signatures and relative phase effects when the standing wave contains a fundamental
frequency and its second harmonic. The influence of the relative field intensities on the Rabi oscillation
dynamics is also discussed.
\end{abstract}

\maketitle

\section{Introduction}

The quantum mechanical diffraction of an electron beam on the periodic potential generated by a standing light wave is referred to as Kapitza-Dirac effect \cite{Kapitza1933,Batelaan2007}.
The process relies on the quantum wave nature of the electron and represents a counterpart of classical diffraction of light on a grating,
with the roles of light and matter interchanged.
In its original version \cite{Kapitza1933} the Kapitza-Dirac effect can be understood as a combined absorption and emission process involving two photons:
The incident electron absorbs one photon of momentum $\hbar\vec{k}$ from one of the laser beams which forms the standing wave,
and emits another photon of momentum $-\hbar\vec{k}$ into the counterpropagating laser beam (stimulated Compton scattering).
The momentum transfer from the standing wave to the electron thus amounts to $2\hbar\vec{k}$.

A clear experimental verification of the Kapitza-Dirac effect as originally proposed has been accomplished only recently.
Utilizing modern state-of-the-art equipment and a laser intensity of the order of $\sim 10^{11}$\,W/cm$^2$,
the expected electron diffraction pattern was recorded \cite{Freimund2001, *Freimund2002}.
Another successful experiment observed electron (rainbow) scattering after atomic ionization in a standing laser wave of higher intensity $\sim 10^{14}$\,W/cm$^2$ \cite{Bucksbaum1988}.
Also beams of atoms can be subject to Kapitza-Dirac scattering \cite{Adams1994, *Freyberger1999}
which was observed both in the so-called Bragg \cite{Gould1986} and diffraction \cite{Martin1988, *Eilzer2014} regimes.

The recent observation of Kapitza-Dirac scattering of electrons has led to a newly revived interest from the theoretical side.
While seminal work on the Kapitza-Dirac effect and its multiphoton generalization to intense laser fields has already been performed in the 1960s \cite{Fedorov1967},
various additional aspects have been studied during the last decade.
The influence of the electronic wavepacket size was examined \cite{Efremov2000}, a generalization to two electrons was proposed \cite{Sancho2010},
and spin effects were investigated \cite{Rosenberg2004, Ahrens2012, *Ahrens2013, Bauke2014}.
Also relativistic treatments of the Kapitza-Dirac effect have been presented based on the Dirac equation \cite{Ahrens2012, *Ahrens2013}
and the Klein-Gordon equation \cite{Haroutunian1975, *Fedorov1980}, respectively.

Interesting features of electron-laser interactions can emerge in two-color light fields containing two different frequency components \cite{Ehlotzky2001}.
In particular, when the frequency ratio is a commensurate number, characteristic quantum interference and relative phase effects may arise.
Two-color effects have been studied for a variety of processes such as laser-assisted electron scattering in atomic potentials \cite{Varro1993, *Kaminski1995, *Milosevic1997},
photoionization \cite{Schafer1992, *Yin1992, *Schumacher1994, *Veniard1995, *Paulus1995}
and high-harmonic generation \cite{Telnov1995, *Bandrauk1997, *MorissonFaria1999, *Milosevic2000, *Kim2005, *Mauritsson2006} from atoms in strong laser fields,
molecular dissociation and chemical reactions \cite{Shapiro2000},
Thomson scattering \cite{Sperling2011, *Sperling2013}, and even electron-positron pair production \cite{Krajewska2012, *Augustin2013}.
Due to the competition of various quantum pathways, these two-color effects offer possibilities to implement coherent control schemes.

Regarding the Kapitza-Dirac effect in bichromatic fields, two different kinds of studies can be found in the literature. 
On the one hand, it was shown that electrons can scatter diffractively from two counterpropagating laser beams even if the latter have non-identical frequencies \cite{Smirnova2004}.
Although no standing wave is formed in this case, the electrons may still experience a periodic effective potential in their rest frame from which they are scattered coherently.
This field configuration has also been examined with respect to electron spin dynamics \cite{Freimund2003} and as an interferometric electron beam splitter \cite{Marzlin2013}.
A bichromatic standing wave, on the other hand, can be formed by using two pairs of counterpropagating laser waves which possess different frequencies $\omega_1$ and $\omega_2$.
Kapitza-Dirac scattering of three-level atoms from this periodic field structure has been investigated \cite{Pazgalev1996}
which offers interesting properties for atomic optics in general \cite{Dubetsky2002, *Giese2013}.
To the best of our knowledge, Kapitza-Dirac diffraction of electrons from a  bichromatic standing light wave has not been considered before.

In this paper we study Kapitza-Dirac scattering of an electron beam from a bichromatic standing light wave with a commensurate frequency ratio.
After establishing a common theoretical framework based on the time-dependent Schr\"odinger equation in Sec.~\ref{sec:theory}., we focus on two distinct parameter regimes.
On the one hand, we investigate in Sec.~\ref{sec:diffraction}. the influence of the bichromaticity of the standing light wave on the electron dynamics in the diffraction regime,
where the field amplitude is high.
Here we present an analytical two-color generalization to the well-known Bessel-like solution for monochromatic waves, which is valid for asymptotically large field amplitudes.
On the other hand, in Sec.~\ref{sec:bragg}. we study Kapitza-Dirac scattering in the Bragg regime where the field amplitudes are perturbatively low and Rabi oscillations between the relevant electron states occur.
A closed-form expression for the Rabi frequency is derived within an approximate analytical model in reduced dimensionality.
In both parameter regimes, we compare the analytical approaches with numerical computations and demonstrate characteristic quantum interference and relative phase effects.
We finish with a conclusion in Sec.~\ref{sec:conclusion}.

\section{Theoretical framework}
\label{sec:theory}
The nonrelativistic domain of electron-light scattering is described by the time-dependent Schr\"odinger equation
\begin{eqnarray}
 i\hbar \frac{\partial}{\partial t} \psi = \frac{1}{2 m} \left( \frac{\hbar}{i} \vec\nabla + \frac{e}{c} \vec{A} \right)^2 \psi
\label{eqn:TD-SE}
\end{eqnarray}
where $\hbar$ is the reduced Planck constant, $m$ the electron mass, $-e$ its charge, $c$ the speed of light and $\psi$ the electron wave function.
Besides,
\begin{eqnarray}
 \vec{A} &=& f(t) A_0 \left[ \alpha \vec{\varepsilon}_1 \cos \left( ckt \right) \cos \left( kz \right) \vphantom{\frac{\sqrt{1-\alpha^2}}{2}} \right. \nonumber\\*
  &+& \left. \frac{\sqrt{1-\alpha^2}}{2} \vec{\varepsilon}_2 \cos \left( 2ckt + \eta \right) \cos\left( 2kz + \frac{\delta}{2} \right) \right]
\label{eqn:vector-potential}
\end{eqnarray}
models the vector potential of a bichromatic standing light wave, which contains a fundamental frequency $\omega_1=kc$ and its second harmonic $\omega_2=2kc$,
with an envelope function $f(t)$, an amplitude factor $A_0$, and two polarization vectors $\vec{\varepsilon}_{1}$ and $\vec{\varepsilon}_{2}$.
The corresponding field amplitudes of the two frequency modes are controlled by the mixing parameter $\alpha$ in such a way,
that the overall peak laser intensity $I=\frac{ck^2A_0^2}{16\pi}$ is independent of $\alpha$.
The relative phases between the two modes in space and time are described by $\delta$ and $\eta$, respectively.
A frequency ratio of $\omega_2/\omega_1=2$ has been chosen because the most pronounced quantum interference effects are to be expected for this case.

From investigation of the Kapitza-Dirac effect in monochromatic waves it is known that, instead of the vector potential $\vec{A}$,
an effective scalar potential may be used to describe the electron-light interaction.
It is obtained by taking a time average over the rapid oscillations of the electron along the polarization direction of the field \cite{Batelaan2007}.
This way, one arrives at the so-called ponderomotive potential
$V(z) = \frac{e^2}{2 m c^2} \left\langle \vec{A}^2 \right\rangle_t$.
In the bichromatic setting, the Schr\"odinger equation \eqref{eqn:TD-SE} in ponderomotive potential approximation becomes
\footnote{Note that, when squaring the vector potential, the spatial cosines produce summands of $\frac{1}{2}f(t)^2$ which can be omitted as spatially constant terms in the ponderomotive potential.}
\begin{eqnarray}
 i\hbar \frac{\partial}{\partial t} \psi &=& -\frac{\hbar^2}{2m} \vec\nabla^2 \psi \nonumber\\*
 &+& f(t)^2 \frac{V_0}{2} \left[ \alpha^2\cos(2kz) + \frac{1-\alpha^2}{4} \cos(4kz + \delta) \right] \psi \nonumber\\*
\label{eqn:2-color-SE}
\end{eqnarray}
where the amplitude of the ponderomotive potential is $V_0=\frac{e^2 A_0^2}{4 m c^2}$.
We note that the relative orientation of the polarization vectors of the two standing waves is immaterial, since the mixed product term vanishes in the averaging process.
The same happens to the temporal relative phase $\eta$.

Having only $z$-dependence in the potential, the latter equation becomes effectively one-dimensional in space.
It can, thus, be solved by an ansatz in the form of an expansion into plane waves
\begin{eqnarray}
  \psi(z, t) = \sum_n i^{-n} c_n\left( \frac{V_0 t}{2 \hbar} \right) e^{\frac{i}{\hbar} \left( 2 n \hbar k + p_z \right) z} \quad ,
\label{eqn:ansatz}
\end{eqnarray}
with time-dependent expansion coefficients $c_n$.
The sum beeing discrete because, due to the periodicity of the potential, only the given discrete subset of momentum eigenstates do interact.
By plugging Eq.~\eqref{eqn:ansatz} into Eq.~\eqref{eqn:2-color-SE} and using the scaled time variable $\tau := V_0 t/(2 \hbar)$, we obtain
\begin{eqnarray}
 \frac{d c_n(\tau)}{d \tau} &=& \frac{\left( p_z + 2 n \hbar k \right)^2}{i m V_0} c_n(\tau) \nonumber\\*
 &+& f(t)^2 \frac{\alpha^2}{2} \left[ c_{n-1}(\tau) - c_{n+1}(\tau) \right] \nonumber\\*
 &+& f(t)^2 \frac{(1-\alpha^2)}{8} \left[ ie^{i\delta} c_{n-2}(\tau) + ie^{-i\delta} c_{n+2}(\tau) \right] \quad . \nonumber\\*
\label{eqn:2-color-plain-wave}
\end{eqnarray}
This coupled system of ordinary differential equations governs the full dynamics of the time-dependent coefficients $c_n$
which represent the occupation amplitudes of the respective plane-wave electron states.\\

One distinguishes two different interaction regimes, depending on the relative magnitude of the ponderomotive potential as compared with the kinetic energy term in Eq.~\eqref{eqn:2-color-plain-wave}:
One refers to the diffraction regime when the former is substantially larger than the latter, whereas to the Bragg regime when the kinetic energy term is dominant.
Kapitza-Dirac scattering exhibits qualitative differences in these regimes.
They will be examined in the subsequent sections.
The initial condition will always be chosen as  $c_n(0) = \delta_{n, 0}$ which describes an incident electron with given longitudinal momentum $p_z$.

\section{Diffraction regime}
\label{sec:diffraction}
From the Kapitza-Dirac effect in monochromatic waves it is known that the Schr\"odinger equation in ponderomotive potential-approximation
can be solved analytically in the asymptotic high-intensity case
\begin{eqnarray}
 \hbar^2 k^2 \ll m f(t)^2 V_0
\label{eqn:diffraction-approximation}
\end{eqnarray}
where the kinetic energy term may be neglected. The scattering amplitudes $c_n$ are then essentially given by ordinary Bessel functions $J_n$ (see Sec. II.B. in \cite{Batelaan2007}).
In the same asymptotic limit, also the Schr\"odinger equation \eqref{eqn:2-color-plain-wave} for Kapitza-Dirac scattering from a bichromatic wave can be solved analytically.
We obtain the solution in the following way:
First let us assume $f \equiv 1$ and introduce the differential operator
\begin{eqnarray}
 \mathcal{D} &:=& \frac{\partial}{\partial \tau} - \frac{\alpha^2}{2} \left( \zeta - \zeta^{-1} \right) \nonumber\\*
  &-& \frac{(1-\alpha^2)}{8} \left( i e^{i\delta} \zeta^2 + i e^{-i\delta} \zeta^{-2} \right) \quad ,
\label{eqn:bessel-operator}
\end{eqnarray}
the number operator
\begin{eqnarray}
 \mathcal{N} &:= \zeta \frac{\partial}{\partial \zeta} \quad,
\label{eqn:number-operator}
\end{eqnarray}
the order parameter $\varepsilon := \frac{4\hbar^2 k^2}{m V_0}$ and the relative momentum offset $p:=\frac{p_z}{\hbar k}$.
With this notation, Eq.~\eqref{eqn:2-color-plain-wave} can be rewritten as a differential equation for the generating function
\begin{eqnarray}
 C(\tau, \zeta) := \sum_{n=-\infty}^{\infty} c_n(\tau) \zeta^n
\label{eqn:generating-function}
\end{eqnarray}
of the coefficients $c_n$ as
\begin{eqnarray}
 \mathcal{D} C(\tau, \zeta) = -i\varepsilon \left( \frac{p}{2} + \mathcal{N} \right)^2 C(\tau, \zeta) \quad.
\label{eqn:generating-function-SE}
\end{eqnarray}
The boundary condition reads $C(0,\zeta) = 1$, which corresponds to $c_n(0)=\delta_{n,0}$.
We note that the generating function formally represents a Laurent series.
In the limit $\varepsilon \to 0$, Eq.~\eqref{eqn:generating-function-SE} is solved exactly by
\begin{eqnarray}
 C^{0}(\tau, \zeta) &=& \exp\left[ \frac{\alpha^2 \tau}{2}\left( \zeta - \zeta^{-1} \right) \right] \times \nonumber\\*
  &\phantom{=}& \exp\left[ \frac{i(1-\alpha^2) \tau}{8}\left( e^{i\delta} \zeta^2 + e^{-i\delta}\zeta^{-2} \right) \right] \nonumber\\*
  &=& J \left( \alpha^2\tau, \zeta \right) J \left( \frac{1-\alpha^2}{4}\tau, i e^{i\delta} \zeta^2 \right)
\label{eqn:bessel-solution}
\end{eqnarray}
where
\begin{eqnarray}
 J(\rho, \zeta) = \sum_{n=-\infty}^{\infty} J_n(\rho) \zeta^n = \exp \left[ \frac{\rho}{2} \left( \zeta - \zeta^{-1} \right) \right]
\label{eqn:bessel-generator}
\end{eqnarray}
is the generating function of ordinary Bessel functions of the first kind.
The solution in Eq.~\eqref{eqn:bessel-solution} serves as a Green's function for $\mathcal{D}$,
and simultaneously can be recognised as the generating function of the generalized Bessel functions $\tilde{J}_n$ \cite{dattoli1991}.
By virtue of Eq.~\eqref{eqn:generating-function}, we can extract the unperturbed coefficients from Eq.~\eqref{eqn:bessel-solution} as
\begin{eqnarray}
 c_n^0(\tau) &=& \tilde{J}_n \left( \alpha^2\tau, \frac{1-\alpha^2}{4}\tau, \delta \right) \nonumber\\*
 &=& \sum_{\ell=-\infty}^{\infty} J_{n-2\ell} \left( \alpha^2 \tau \right) i^{\ell}e^{i\ell\delta} J_{\ell} \left( \frac{1-\alpha^2}{4} \tau \right) \,.
\label{eqn:generalized-bessel}
\end{eqnarray}
One can see, that this approximate solution depends on $V_0$ and $t$ only through their product.
This scaling law implies that Eq.~\eqref{eqn:generalized-bessel} can be recast to cover the more general case of a switching function $f$ different from unity by replacing the action parameter $\tau$ with
\begin{eqnarray}
 \tilde{\tau} = \frac{V_0}{2 \hbar} \int_{0}^{t} f(t')^2\, dt' \quad.
\label{eqn:tau-action}
\end{eqnarray}
It is assumed here that the part of the integral which does not satisfy the condition \eqref{eqn:diffraction-approximation} is negligible.
This assumption corresponds to fast switching of the electron-light interaction.

\begin{figure}[b]
\begin{center}
 \includegraphics{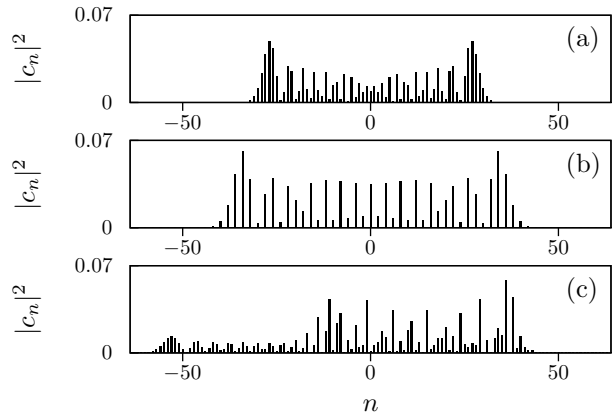}
\end{center}
\caption{Longitudinal momentum distributions of electrons after Kapitza-Dirac diffraction from
 (a) a monochromatic standing wave with $V_1=3.6 \times 10^{-2}\,\mathrm{eV}$ and $k_1=2\,\mathrm{eV}/\hbar c$,
 (b) a monochromatic standing wave with $V_2=2.7 \times 10^{-2}\,\mathrm{eV}$ and $k_2=4\,\mathrm{eV}/\hbar c$,
 and (c) a bichromatic standing wave with $V_1=3.6 \times 10^{-2}\,\mathrm{eV}$, $k_1=2\,\mathrm{eV}/\hbar c$,
 $V_2=2.7 \times 10^{-2}\,\mathrm{eV}$, $k_2=4\,\mathrm{eV}/\hbar c$, and $\delta=\frac{\pi}{2}$.
 The interaction time is $t = 1.1 \times 10^{-12}\,\mathrm{s}$.}
\label{fig:bessel}
\end{figure}

From the asymptotic solution \eqref{eqn:bessel-solution} we can already derive an interesting property of Kapitza-Dirac diffraction in the bichromatic case.
Since the two field frequencies are commensurate ($\omega_2 = 2\omega_1$), quantum mechanical interferences may arise.
It is indistinguishable whether, for example, two photons of low frequency or one photon of high frequency have been absorbed
since their (total) energy and momentum are identical.
This feature is illustrated in Fig.~\ref{fig:bessel} where the Kapitza-Dirac scattering pattern is shown for each monochromatic wave alone [panels (a) and (b)]
and for the combined, bichromatic wave [panel (c)].
It can be clearly seen that the bichromatic case is not obtained as a simple summation of the monochromatic cases, demonstrating the presence of quantum interference.
In particular, while the monochromatic patterns are symmetric with respect to the transformation $n \to -n$,
the bichromatic pattern shows a characteristic asymmetry which depends on the relative phase between the frequency modes.

\begin{figure}[t]
\begin{center}
 \includegraphics{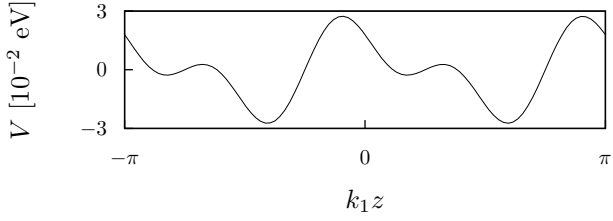}
\end{center}
\caption{Ponderomotive potential $V$ of a bichromatic standing laser wave for the field parameters of Fig.~\ref{fig:bessel}(c).}
\label{fig:field}
\end{figure}

The basic reason for this feature is that the standing wave itself can be asymmetric when it contains two frequency components (see Fig.~\ref{fig:field}).
Therefore, it may be more likely that the incident electron beam is scattered to the right than to the left (or vice versa).
Only for $\delta = 0$ or $\pi$, both the standing wave and the scattering pattern turn out to be symmetric again, as one may have expected.

The question arises how good an approximation Eq.~\eqref{eqn:bessel-solution} is or, in other words,
how long the interaction time may last at most until the kinetic energy term causes significant deviations of the exact (numerical) solution from the approximation \eqref{eqn:bessel-solution}.
To answer this question, we consider finite values of the small parameter $\varepsilon = \frac{4 \hbar^2 k^2}{m V_0}$
and calculate corrections to Eq.~\eqref{eqn:bessel-solution} by using time-dependent perturbation theory.
With the expansion ansatz
\begin{eqnarray}
 C(\tau, \zeta) = C^0(\tau, \zeta) + \varepsilon C^1(\tau, \zeta) + \varepsilon^2 C^2(\tau, \zeta) + \dots
\label{eqn:C-expansion-ansatz}
\end{eqnarray}
Eq.~\eqref{eqn:generating-function-SE} decomposes into a Dyson series. To the first order in $\varepsilon$, we obtain
\begin{eqnarray}
 C^1(\tau, \zeta) &=& -i \int_0^\tau d\sigma C^0(\tau-\sigma, \zeta) \left( \frac{p}{2} + \mathcal{N} \right)^2 C^0(\sigma, \zeta) \nonumber\\*
 &=& -i \left( \frac{\tau}{4} p^2 + \frac{\tau}{2} p \mathcal{N} + \frac{\tau}{3} \mathcal{N}^2 + \frac{\tau^2}{6} [\mathcal{N}, [\mathcal{N}, \mathcal{D}]] \right) \nonumber\\*
 &\times& C^0(\tau, \zeta)
\label{eqn:first-order}
\end{eqnarray}
where $[,]$ indicates a commutator. This means
\begin{eqnarray}
 c_n(\tau) &=& \left( 1 - i\varepsilon \frac{\tau}{4} p^2 - i\varepsilon \frac{\tau}{2} n p - i\varepsilon \frac{\tau}{3} n^2 \right) c_n^0(\tau) \nonumber\\*
 &-& i\varepsilon \frac{\alpha^2 \tau^2}{12} \left[ c_{n-1}^0(\tau) - c_{n+1}^0(\tau) \right] \nonumber\\*
 &+& \varepsilon \frac{(1-\alpha^2) \tau^2}{12} \left[ e^{i\delta} c_{n-2}^0(\tau) + e^{-i\delta} c_{n+2}^0(\tau) \right] + o(\varepsilon) \nonumber\\*
\label{eqn:first-order-n}
\end{eqnarray}
With the Taylor expansions
\begin{eqnarray}
 c_{0}^0(\tau) &=& 1 - \left[ \alpha^2 + \frac{\left( 1-\alpha^2 \right)^2}{16} \right] \left( \frac{\tau}{2} \right)^2 + o\left( \tau^2 \right) \nonumber \\*
 c_{\pm 1}^0(\tau) &=& \pm \alpha^2 \frac{\tau}{2} \mp i e^{\pm i\delta} \frac{ \alpha^2 \left( 1-\alpha^2 \right)}{4} \left( \frac{\tau}{2} \right)^2 + o\left( \tau^2 \right) \nonumber \\*
 c_{\pm 2}^0(\tau) &=& i e^{\pm i \delta} \frac{1-\alpha^2}{4} \frac{\tau}{2} + \frac{1}{2} \left( \frac{\tau}{2} \right)^2 + o\left( \tau^2 \right)
\label{eqn:bessel-taylor}
\end{eqnarray}
we find that
\begin{eqnarray}
 \frac{\left\vert c_0(\tau) \right\vert^2}{\left\vert c_0^0(\tau) \right\vert^2}
 = 1 + \varepsilon \tau^4 \frac{1}{4}\alpha^4 \frac{1-\alpha^2}{4} \cos \delta + o\left( \varepsilon \tau^4 \right)
\label{eqn:c0-relative-error}
\end{eqnarray}

\begin{figure}[h]
\begin{center}
\includegraphics[]{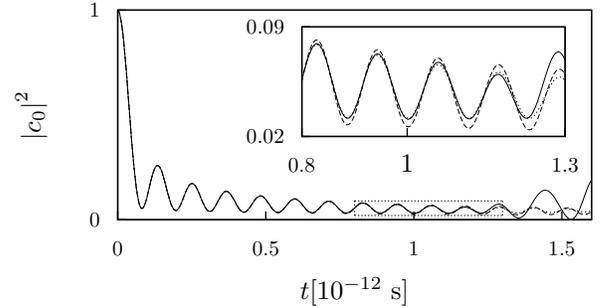}
\end{center}
\caption{Time evolution of the forward scattering probability $|c_0|^2$ evaluated by the asymptotic formula \eqref{eqn:bessel-solution} [dashed line],
  the first-order corrected Eq.~\eqref{eqn:first-order} [dotted line], and the exact numerical solution of Eq.~\eqref{eqn:2-color-SE} [solid line].
  The bichromatic laser intensity is $I = 10^{12}\,\mathrm{W/cm^2}$, its fundamental wave number $k=2\,\mathrm{eV}/\hbar$ and it is mixed with the second harmonic
  by the parameters $\alpha^2 = 0.5$, $\delta = 0$.
  The inset shows a magnification of the marked area.}
\label{fig:comparison}
\end{figure}

Figure~\ref{fig:comparison} shows the squared modulus of the initial momentum mode as a function of the interaction time.
The predictions from Eq.~\eqref{eqn:bessel-solution}, the first-order corrected Eq.~\eqref{eqn:first-order}, and an exact numerical computation are compared.
We can see that the first-order approximation follows the numerical solution quite nicely and for a longer time than the asymptotic approximation by Bessel functions.
This is to be expected from Eq.~\eqref{eqn:c0-relative-error} for the specific choice of parameters $\alpha^2 = 0.5$ and $\delta = 0$.
The order parameter, in this example, is $\varepsilon = 4.4 \times 10^{-4}$ and the time $t=10^{-12}\,\mathrm{s}$ corresponds to $\tau \approx 55$.
Thus, the analytical expressions provide good approximations for even much longer interaction times than one might have expected from Eq.~\eqref{eqn:c0-relative-error}.

\section{Bragg regime}
\label{sec:bragg}
In this section we consider Kapitza-Dirac scattering in the case of a smooth, slowly varying switching-function $f(t)$ and rather small ponderomotive potential $V_0$.
Here the adiabatic theorem \cite{Born1928,Fedorov1974} suggests, that the incident electron can only be scattered into a final state with the same kinetic energy.
Due to the fact that only those electrons with specific angles of incidence satisfying a Bragg condition can undergo the transition into the mirrored momentum state,
this parameter range ist called Bragg regime.

Since the transition from one momentum state to its mirrored state is working either ways, Rabi-like oscillations between both states may occur.
This kind of Rabi oscillations is well understood in the Bragg regime of the classical single-color Kapitza-Dirac effect (called ``Pendell\"{o}sung'' in \cite{Batelaan2007}).
The two states of equal energy are resonantly coupled by the interaction with the standing light wave,
causing the electron population probability to oscillate forth and back between them.
Our goal is to shed some light on the influence exerted by the presence of the second laser mode on the Rabi oscillation dynamics.
As before, the frequency of the second mode is assumed to be twice as large as the frequency of the fundamental mode.
We note that in this setup, a Bragg condition can be satisfied by certain incident angles for both laser modes simultaneously.
The first such case is described by an incident longitudinal momentum $p_z=-2\hbar k$ where the initial state and its associated mirrored state are given by $c_{0}$ and $c_{2}$ [see Eq.~\eqref{eqn:ansatz}] which are considered in the following.

\begin{figure}
  \includegraphics{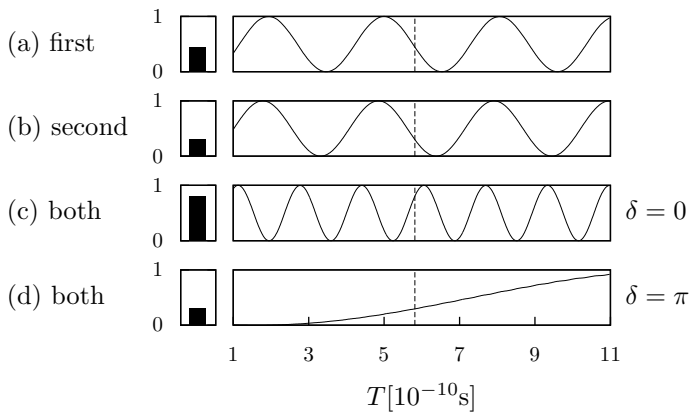}
  \caption{Scattering probabilities $\vert c_2 \left( \frac{V_0 T}{2\hbar} \right) \vert^2$
    after $T = 581\,\mathrm{ps}$ effective interaction time (left)
    and Rabi oscillation dynamics (right) for Kapitza-Dirac scattering from
    (a) a monochromatic standing wave with $k_1=4\,\mathrm{eV}/\hbar c$ and
    $I_1=5.0 \times 10^{9}\,\mathrm{W/cm^2}$, (b) a monochromatic standing
    wave with $k_2=8\,\mathrm{eV}/\hbar c$ and $I_2=6.0\times
    10^{9}\,\mathrm{W/cm^2}$, and (c) [(d)] a bichromatic standing wave with
    the combined parameters and $\delta=0$ [$\delta=\pi$]. The mixing parameter
    in the bichromatic case amounts to $\alpha^2=0.45$,
    the ponderomotive potentials to $V_1 = 9.0\times 10^{-5}\,\mathrm{eV}$ and $V_2 = 2.8 \times 10^{-5}\,\mathrm{eV}$.
    The electron is always incident with a longitudinal momentum of $-8\,\mathrm{eV}/c$,
    and its transverse momentum can be chosen arbitrarily.
    Every plotted data corresponds to a full interaction with switching on and off of $10^{-10}\,\mathrm{s}$ duration each.}
  \label{fig:rabi}
\end{figure}

In Fig.~\ref{fig:rabi} we show the scattering probability (left) and the Rabi-like dynamics of Kapitza-Dirac scattering for various field configurations.
The upper two panels (a) and (b) refer to monochromatic standing waves, with photon energies of $\hbar\omega_1 = 4$\,eV and $\hbar\omega_2 = 8$\,eV, respectively.
Their ponderomotive amplitudes are tuned to give the same Rabi frequency by adjusting the field intensities appropriately. 
The lower two panels (c) and (d) show the results for bichromatic fields,
which are formed by superimposing the two monochromatic standing waves with a relative phase shift of zero and $\pi$, respectively.
The envelope function is chosen such that $f(t)^2$ is flat top of magnitude unity with $\sin^2$ edges of duration $100\,\mathrm{ps}$ each.
The effective interaction time $T$ is defined as the temporal integral of $f(t)^2$.
As the scattering probabilities into the state $c_2$ after $T = 581\,\mathrm{ps}$ show,
the influence of two-color quantum interferences is substantial.
Depending on the chosen relative phase, they can be constructive ($\delta=0$) or destructive ($\delta=\pi$).
The temporal evolution of the scattering probabilities shows besides that the Rabi frequency strongly depends on the relative phase as well.
For a vanishing relative phase, the Rabi frequency is about twice as large as in the monochromatic cases, whereas for $\delta=\pi$, it is heavily suppressed by an order of magnitude.
Note that the effective time in the graphs starts at $100\,\mathrm{ps}$ which corresponds to one full switching cycle and vanishing plateau time.

\begin{figure}
  \includegraphics{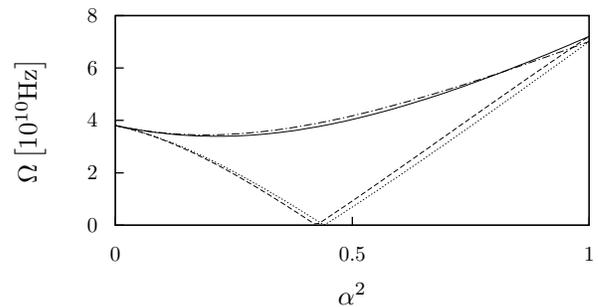}
  \caption{Rabi frequency $\Omega$ of the bichromatic Kapitza-Dirac scattering process in the Bragg regime, as a function of the mixing parameter $\alpha^2$.
  The wave numbers being $k_1 = 4\,\mathrm{eV} / \hbar c$, $k_2 = 8\,\mathrm{eV}/\hbar c$, and the combined intensity $I = 1.1 \times 10^{10}\,\mathrm{W/cm^2}$.
  The relative phase is $\delta = 0$ for the solid line and $\delta = \pi$ for the dashed line.
  The dash-dotted and dotted lines result from the approximate analytical formula \eqref{eqn:omega_rabi}, with the same parameters.}
  \label{fig:rabi_alpha}
\end{figure}
\begin{figure}
  \includegraphics{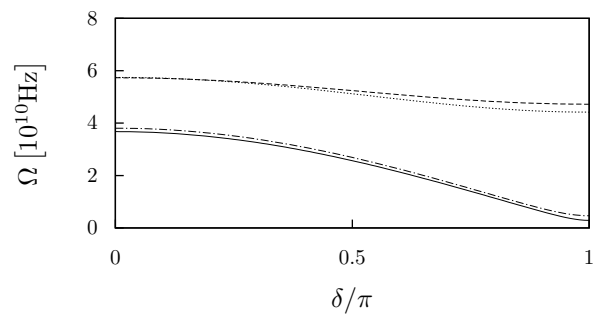}
  \caption{Rabi frequency $\Omega$ of the bichromatic Kapitza-Dirac scattering process in the Bragg regime, as a function of the relative phase $\delta$.
  The wave numbers and combined intensity are as in Fig.~\ref{fig:rabi_alpha}.
  The mixing parameter is $\alpha^2 = 0.4$ for the solid line and $\alpha^2 = 0.8$ for the dashed line.
  The dash-dotted and dotted lines result from the approximate analytical formula \eqref{eqn:omega_rabi}, with the same parameters.}
  \label{fig:rabi_delta}
\end{figure}

To further investigate this dependence Fig.~\ref{fig:rabi_alpha} shows the Rabi frequency $\Omega$ in a bichromatic standing wave of fixed intensity,
as a function of the mixing parameter $\alpha^2$ for relative phases of $\delta = 0$ and $\pi$, respectively.
According to Eq.~\eqref{eqn:2-color-SE}, a mixing parameter of $\alpha^2=1$ corresponds to a monochromatic standing wave with the fundamental frequency $\omega_1$,
whereas $\alpha^2=0$ corresponds to a monochromatic wave with the second harmonic frequency $\omega_2$.
In these limiting cases, the relative phase becomes immaterial so that the solid and dashed lines coincide here.
However, when both frequency modes are present (i.e., for $0<\alpha^2<1$), the impact of the relative phase is pronounced.
For $\delta=0$, the Rabi frequency depends on the color mixing only moderately:
starting from $\alpha^2=0$, it smoothly passes through a shallow minimum and afterwards raises to reach somewhat higher values.
We point out that the Rabi frequency for $\alpha^2=1$ is larger than for $\alpha^2=0$ because the laser waves have a fixed combined intensity along the curve [cp. Figs.~\ref{fig:rabi}a) and b) where $I_2>I_1$ instead].
Contrary to that, for $\delta=\pi$, the Rabi frequency shows a strong dependence on the mixing parameter:
for bichromatic fields with roughly equal intensity shares between both modes, it drops down heavily and approaches zero for $\alpha^2\approx 0.4$.

Similarly, Fig.~\ref{fig:rabi_delta} illustrates the dependence of the Rabi frequency on the relative phase $\delta$ for two values of the mixing parameter.
For $\alpha^2=0.4$, the Rabi frequency is very sensitive to the value of $\delta$ and decreases by an order of magnitude when $\delta$ varies from 0 to $\pi$.
Instead, for $\alpha^2=0.8$, when the admixture of the second color is only small, the Rabi frequency varies only slightly with the relative phase.
It is interesting to note that a similar behavior of the Rabi frequency like in Figs.~\ref{fig:rabi_alpha} and \ref{fig:rabi_delta} can also be seen for other commensurate frequency ratios, as we have checked by corresponding calculations.

The dependencies of the Rabi frequency on the various parameters of a bichromatic standing light wave can roughly be understood within a model of greatly reduced dimensionality.
Let us consider only the three electron momentum modes with $n\in\{0,1,2\}$ and combine them into a tupel $u:=(c_0, -i c_{1}, -c_{2}) \in \mathbb{C}^3$ for $p_z=-2\hbar k$ and $f(t)^2\equiv 1$.
Then, Eq.~\eqref{eqn:2-color-plain-wave} can be rewritten as
\begin{eqnarray}
 i \frac{V_0}{2} \frac{du}{d\tau} = M u \quad,
\label{eqn:SG-reduced-dim}
\end{eqnarray}
where we have introduced the hermitian matrix
\begin{eqnarray}
M=\left( \begin{matrix} A & B & D \\ B & 0 & B \\ D^* & B & A \end{matrix} \right)
\end{eqnarray}
with entries $A=\frac{2\hbar^2 k^2}{m}$, $B=V_0 \frac{\alpha^2}{4}$ and $D=V_0 \frac{1-\alpha^2}{16}e^{-i\delta}$.
The characteristic polyomial of $M$ reads
\begin{eqnarray}
 \chi(\lambda) &=& \lambda^3 - 2A \lambda^2 + \left( A^2 - 2B^2 - \vert D \vert^2 \right) \lambda \nonumber\\*
 &+& 2B^2\left( A- \Re D \right).
\label{eqn:cpm}
\end{eqnarray}
The roots of this cubic polynomial, which are the eigen\-energies of the system in Eq.~\eqref{eqn:SG-reduced-dim}, are given by
\begin{eqnarray}
 \lambda_j &=& \frac{2}{3}A + \sqrt{\frac{4}{9}A^2 + \frac{8}{3}B^2 + \frac{4}{3} \vert D \vert^2} \cos \left\{ \frac{1}{3} \right. \nonumber\\*
 &&\times \arccos \left[ \left( -\frac{1}{27}A^3 - \frac{1}{3}AB^2 + \frac{1}{3}A \vert D \vert^2 + B^2 \Re D \right) \right. \nonumber\\*
 &&\times \left. \left. \left( \frac{1}{9} A^2 + \frac{2}{3} B^2 + \frac{1}{3} \vert D \vert^2 \right)^{-\frac{3}{2}} \right] + \frac{2}{3}j \pi \right\}
\label{eqn:roots}
\end{eqnarray}
for $j\in\{-1, 0, 1\}$.
Taking the limit $B, D \to 0$, which models smoothly switching off the interaction, the two eigenvalues for $j = -1$ and $j=0$ tend to $A$.
\begin{figure}[h]
  \includegraphics{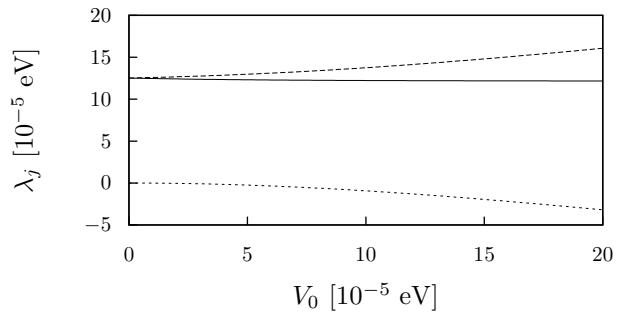}
  \caption{Roots $\lambda_{j}$ of the polynomial \eqref{eqn:cpm} as given in Eq.~\eqref{eqn:roots}, as a function of the ponderomotive potential $V_0$.
  The wavenumbers are $k_1=4\,\mathrm{eV}/\hbar c$ and $k_2=8\,\mathrm{eV}/\hbar c$, the mixing parameter $\alpha^2=\frac{1}{2}$, and the relative phase $\delta=\frac{\pi}{2}$.
  The solid, dashed, and dotted lines refer to $j=-1$, $j=0$, and $j=1$, respectively.}
  \label{fig:roots}
\end{figure}
This is also illustrated in Fig.~\ref{fig:roots}.
The corresponding eigenvectors, in that limit, are $v_{-1} = (1, 0, e^{i\delta})$ and $v_0 = (1, 0, -e^{i\delta})$ which span a twofold degenerated subspace.
Our initial condition corresponds to the tupel
\begin{equation}
  u = (1,0,0) = \frac{1}{2} \left( v_{0} + v_{-1} \right)
\label{eqn:inicond}
\end{equation}
which can be expressed as a linear combination of $v_{-1}$ and $v_0$.
The Rabi frequency therefore arises as a beat frequency between the associated eigenfrequencies and is given by

\begin{eqnarray}
 \Omega &=& \frac{1}{\hbar}\left( \lambda_{0} - \lambda_{-1} \right) \nonumber\\*
 &=& \frac{1}{\hbar} \sqrt{\frac{16 \hbar^4 k^4}{9 m^2} + \frac{1}{6} V_0^2 \alpha^4 + \frac{1}{192} V_0^2 \left( 1-\alpha^2 \right)^2} \nonumber\\*
 &&\times \left[ \cos \Phi - \cos \left( \Phi - \frac{2 \pi}{3} \right)\right] ,
\label{eqn:omega_rabi}
\end{eqnarray}
with
\begin{eqnarray}
 \Phi &=& \frac{1}{3} \arccos \nonumber\\*
 && \left\{
   \frac{-\frac{8 \hbar^6 k^6}{27 m^3}
   + \frac{\hbar^2 k^2 V_0^2 \left( \left( 1-\alpha^2 \right)^2 - 16 \alpha^2 \right)}{384 m}
   + \frac{V_0^3 \alpha^4 \left( 1-\alpha^2 \right) \cos(\delta)}{256}}
   {\left[\frac{4 \hbar^4 k^4}{9m^2} + \frac{1}{24} V_0^2\alpha^4 + \frac{1}{768} V_0^2 \left( 1-\alpha^2 \right)^2\right]^{3/2}}
 \right\} \quad.
\end{eqnarray}
After smoothly switching off the interaction, the quantum electron state returns to the aforementioned twodimensional subspace and is consecutively measured in the momentum base $c_n$.
It follows, that only the energy conserving momentum states $c_{0}$ and $c_{2}$, that span the same subspace, interact with each other.

The shape of the graphs in Figs.~\ref{fig:rabi_alpha} and \ref{fig:rabi_delta} are qualitatively well described by Eq.~\eqref{eqn:omega_rabi} including the case of nearly vanishing interaction.
The remaining quantitative differences are due to the fact, that the model restricts the interaction to only one intermediate state,
whereas the original Eq.~\eqref{eqn:2-color-plain-wave} allows for contributions of all possible paths through the accessible momentum states.

We emphasize that the Rabi frequency in Eq.~\eqref{eqn:omega_rabi} vanishes exactly if $\delta=\pi$ and the mixing parameter fulfills the condition
\begin{equation}
 \alpha^2 = \frac{4 \sqrt{1 + 32 \frac{\hbar^2 k^2}{m V_0} + 16 \frac{\hbar^4 k^4}{m^2 V_0^2}} - 16 \frac{\hbar^2 k^2}{m V_0} - 1}{15} \quad.
\label{eqn:alpha_vanish}
\end{equation}
The interaction can therefore be fully suppressed by destructive interference in the analytical model.
While the results of our numerical simulations induce small modifications to Eq.~\eqref{eqn:alpha_vanish},
they are, within the numerical accuracy, consistent with this general conclusion.

\section{Conclusion}
\label{sec:conclusion}
Kapitza-Dirac scattering of electrons from a bichromatic standing wave was considered.
Fokussing on a commensurate frequency ratio of two, we demonstrated distinct quantum interference and relative phase effects in the scattered electron momentum distribution.

In the diffraction regime, an analytical formula for the scattering amplitudes in terms of generalized Bessel functions was derived for asymptotically large ponderomotive potentials.
The range of applicability of this formula was quantified by comparisons with a first-order corrected expression and fully numerical results.
Quantum interferences may lead here to a characteristic asymmetry of the diffraction pattern.

In the Bragg regime of two-color Kapitza-Dirac scattering, we obtained an analytical formula for the Rabi frequency which determines the population dynamics between the relevant electron states.
The formula shows very good agreement with our numerical calculations covering the momentum space in full dimensionality.
We have demonstrated that, by a suitable choice of intensity ratio and relative phase in the bichromatic wave,
the Rabi frequency can be substantially enhanced or strongly (even totally) suppressed.

\section*{Acknowledgement}
This study was supported by SFB TR18 of the German Research Foundation (DFG) under project No. B11.

%

\end{document}